\documentclass[prb,aps,amsfonts,amssymb,superscriptaddress,twocolumn]{revtex4-1}
\usepackage{graphicx}
\usepackage{color}
\usepackage{amsmath, amsthm, amssymb}
\usepackage{float}
\usepackage{sidecap}

\begin{document}
\bibliographystyle{apsrev}
\title{Aging and reduced bulk conductance in thin films of the topological insulator Bi$_2$Se$_3$}



\author{R. Vald\'es Aguilar}
\email{rvaldes@pha.jhu.edu}
\affiliation{The Institute for Quantum Matter, Department of Physics and Astronomy, The Johns Hopkins University, Baltimore, MD 21218 USA.}
\author{L. Wu}
\affiliation{The Institute for Quantum Matter, Department of Physics and Astronomy, The Johns Hopkins University, Baltimore, MD 21218 USA.}
\author{A.V. Stier}
\affiliation{The Institute for Quantum Matter, Department of Physics and Astronomy, The Johns Hopkins University, Baltimore, MD 21218 USA.}
\author{L.S. Bilbro}
\affiliation{The Institute for Quantum Matter, Department of Physics and Astronomy, The Johns Hopkins University, Baltimore, MD 21218 USA.}
\author{M. Brahlek}
\affiliation{Department of Physics and Astronomy, Rutgers the State University of New Jersey. Piscataway, NJ 08854}
\author{N. Bansal}
\affiliation{Department of Physics and Astronomy, Rutgers the State University of New Jersey. Piscataway, NJ 08854}
\author{S. Oh}
\affiliation{Department of Physics and Astronomy, Rutgers the State University of New Jersey. Piscataway, NJ 08854}
 \author{N.P. Armitage}
 \email{npa@pha.jhu.edu}
 \affiliation{The Institute for Quantum Matter, Department of Physics and Astronomy, The Johns Hopkins University, Baltimore, MD 21218 USA.}
 
\begin{abstract}
We report on the effect of exposure to atmospheric conditions on the THz conductivity of thin films of the topological insulator Bi$_2$Se$_3$. We find: 1) two contributions of mobile charge carriers to the THz conductivity immediately after growth, and 2) the spectral weight of the smaller of these decays significantly over a period of several days as the film is exposed to ambient conditions, while the other remains relatively constant. We associate the former with a bulk response, and the latter with the surface. The surface response exhibits the expected robustness of the carriers from 2D topological surface states. We find no evidence for a third spectral feature derived from topologically trivial surface states.
\end{abstract} 

\maketitle

One of the outstanding issues in the field of topological insulators \cite{Hasan-Kane-10,Hasan-Moore-10,Qi-Zhang-11} has been the unique identification of the topological metallic surface states in transport measurements made on both single crystals and thin films. Signatures of the surface response have been contaminated by appreciable bulk conduction. Only recently have clear signatures of 2D behavior been identified in transport experiments \cite{Qu10a, Butch10a,Chen:11a,Xiong:2011fk,Bansal11a,Steinberg:11a}. In particular, 2D behavior has been demonstrated using thin films of Bi$_2$Se$_3$ of different thicknesses and studying the thickness dependence of the transport properties \cite{Bansal11a}. This study showed that a contribution to the conductance was independent of the film's thickness, and thus 2D in origin. The bulk carriers dominated the conductance response for film thicknesses larger than 300 nm. These same films have been used to study the Kerr rotation induced by magnetic field from the surface states \cite{Rolando-Kerr}. Even though the films were exposed for several days to atmospheric conditions, the Kerr response changed little, indicating the robustness of these states against aging due to exposure to atmospheric conditions.

Recently, experiments that explored the effects of experimental conditions on the response of the surface and bulk \cite{Brahlek-11a} have been reported. \citet{Bianchi10} studied the co-existence of trivial (i.e. without topological properties) and topological surface states in single crystals of Bi$_2$Se$_3$ using photoemission experiments. During these experiments the sample was kept in ultrahigh vacuum, and it was argued that the trivial states are formed due to quantum confinement of the 3D electronic structure near the surface, due to so-called band-bending. In addition, \citet{King11} showed that the largest Fermi momentum crossing of the surface states, and accordingly the total carrier density, changed with time on a timescale of a few hours, even as the sample was kept in ultrahigh vacuum. They argued that this change was due to adsorption of residual gases present in the ultrahigh vacuum chamber, such as CO, on the surface. Similar results were obtained by deliberate dosing the surface of freshly cleaved single crystals with water molecules by \citet{Benia11}. In this work, it was explicitly demonstrated that upon further water exposure, the Fermi energy increased up to values of $\approx$ 0.6 eV, while at the same time topologically trivial surface states were induced due to confinement. These studies raise the question of what the effect is on the transport properties of topological insulators of exposing the them to atmospheric gases.

In this letter we explore the effects of exposure to atmospheric conditions on the surface states in thin films of the topological insulator Bi$_2$Se$_3$. We use time domain terahertz spectroscopy (TDTS) to study the finite frequency properties of thin films of Bi$_2$Se$_3$ grown by molecular beam epitaxy on sapphire (Al$_2$O$_3$) (0001) substrates. By using a spectroscopic probe, we are able to separate the effects of aging on the scattering rate (mobility) and on the spectral weight (the ratio of carrier concentration to the effective mass). We report data taken on 2 thin films grown at Rutgers University as reported elsewhere \cite{Bansal11a}. The data were taken approximately 30 minutes after the samples, originally sealed in vacuum, were received at JHU. The samples were mounted on a cold finger of a continuous He gas flow cryostat and cooled to 4 K in one hour. The thin films remained in He atmosphere for the remainder of the low temperature experiment.  All data discussed in this letter were taken at a temperature below 7 K. After the low temperature experiment, the samples were warmed up to room temperature in 5 hours, and then taken out of the cryostat and left exposed to atmospheric conditions for distinct time intervals, typically at least 24 hrs.

\begin{SCfigure*}
\centering
\includegraphics[width=.75\textwidth]{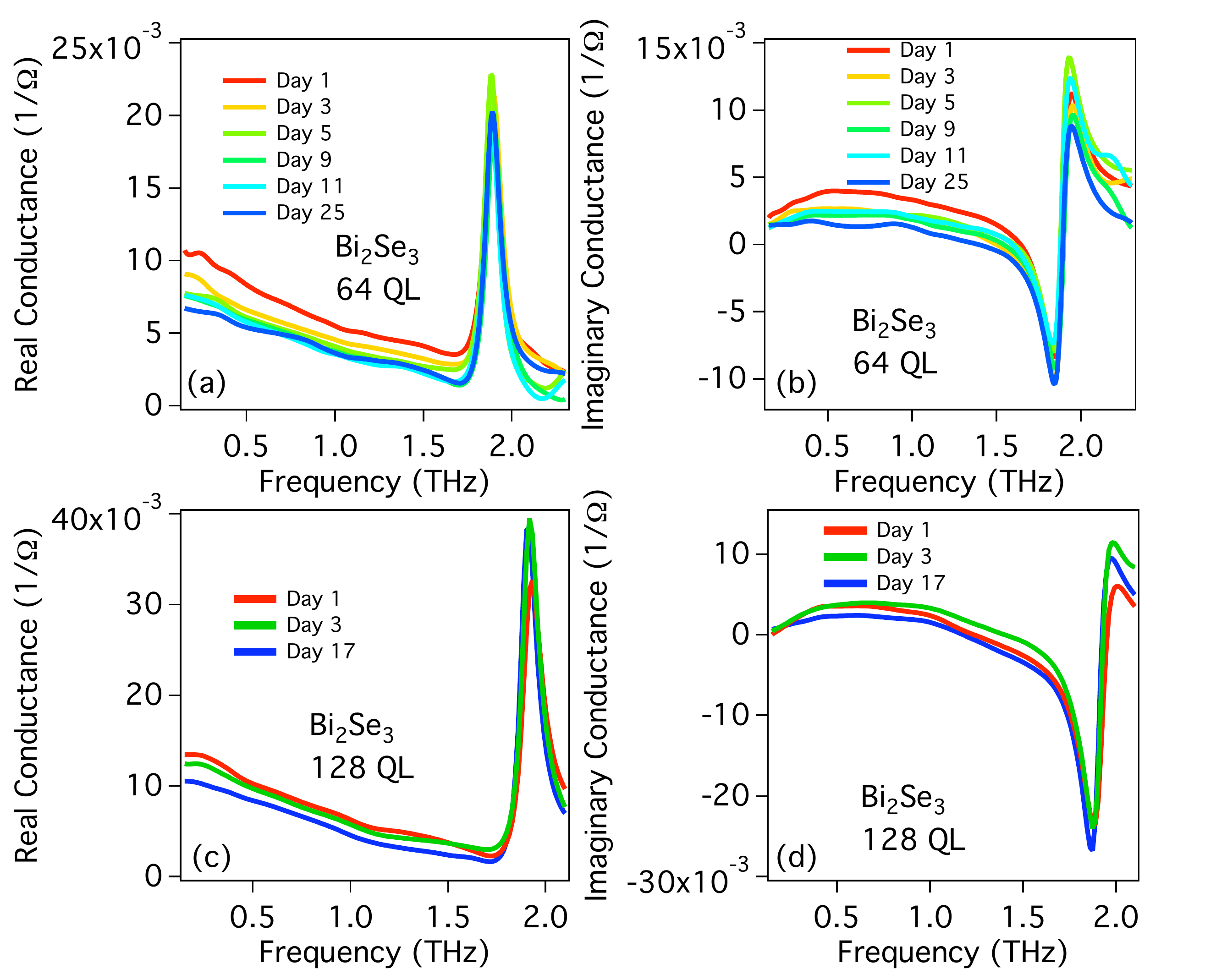}
\caption{(Color Online). \textbf{(a)} Real and \textbf{(b)} imaginary parts of the conductance of a thin film of 64 QL (1 QL=9.4\AA) measured on different days after growth. \textbf{(c)} and \textbf{(d)} same as \textbf{(a)} and \textbf{(b)} for a 128 QL film. It is clear from \textbf{(a)} and \textbf{(c)} that a dramatic decrease on the Drude spectral weight is observed as the film is further exposed to atmospheric conditions.}
\label{Fig1}
\end{SCfigure*}

The terahertz (THz) spectra were obtained in a home-built TDTS system at JHU between frequencies of 0.15 to 2.50 THz. The THz beam was generated by exciting photo-carriers in a DC-biased Auston switch antenna with a pulsed infrared laser with pulses of approximately 50 fs of width. The antenna emits radiation with a bandwidth of a few THz. The THz beam is collimated and then transmitted through the sample using a series of lenses and mirrors, and finally collected and focused onto another antenna where detection of the THz wave occurs. The detecting antenna is also excited by the pulsed infrared beam and works analogously to the emitter, but in this case the photo-carriers are accelerated by the impinging THz wave rather than by a DC voltage. The current generated by the photo-carriers is proportional to the electric field of the THz beam. As the antenna only works while it is illuminated by the ultrashort infrared pulse, one can sample the THz waveform in time by changing the relative time delay between the exciting pulses at the emitter and detector. This ability to measure the electric field as a function of time allows this technique to measure the full optical response, in the form of the complex conductivity, without the use of Kramers-Kronig transformation.

\begin{figure*}[t]
\includegraphics[width=2.1\columnwidth]{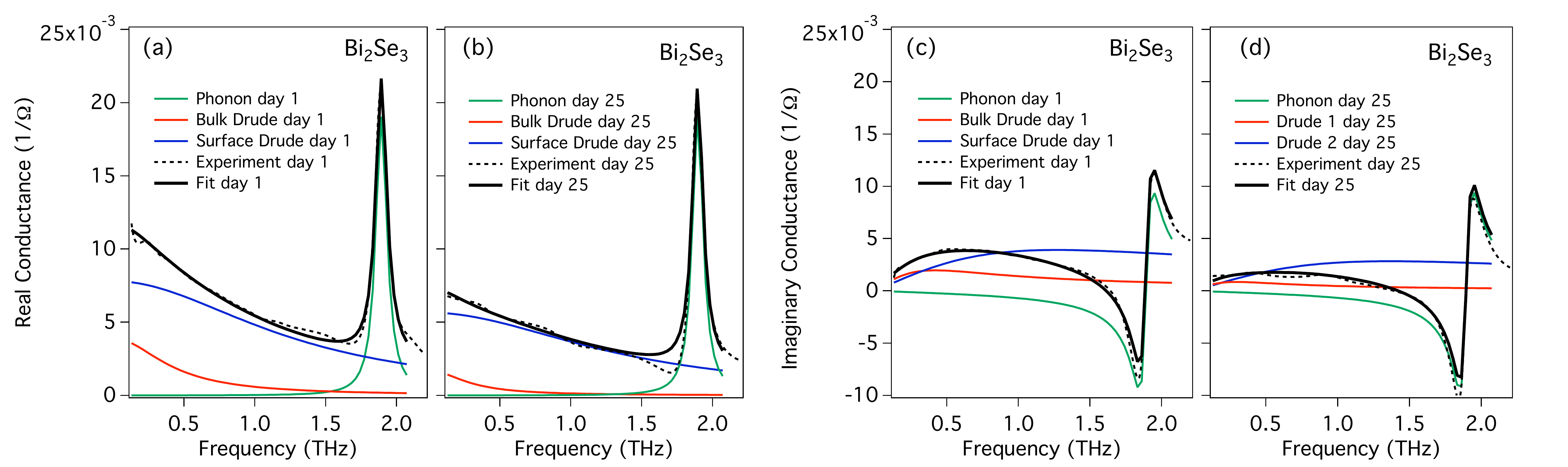}
\caption{(Color Online). \textbf{(a)} and \textbf{(b)}. Real part of the conductance of a thin film of 64 QL in days 1 and 25, respectively, as a function of frequency. The different contributions to the conductance are plotted as lines of different colors. \textbf{(c)} and \textbf{(d)}. Imaginary part of the conductance of the same sample on the same days.}
\label{Fig2}
\end{figure*}

Figure \ref{Fig1} shows the data from a 64 QL sample measured over a span of 4 weeks, and for a 128 QL film measured for almost 3 weeks. Panel (a) shows the real part of the conductance for the 64 QL film, which is composed of a free-electron-like (Drude) response at the lowest frequencies and an absorption peak due to an optical phonon around 2 THz. In our previous work \cite{Rolando-Kerr} we found that, for samples that were kept at atmospheric conditions for at least 2 weeks, the Drude response did not depend on thickness, and we assigned it as a 2D response. For frequencies below the phonon peak, we find that if one tracks the conductivity of a newly grown sample with time, the conductance decreases with time as shown in panel (a) and (c). At the same time, it is clear that the scattering rate (half-width at half maximum) of the Drude response is not affected significantly. This behavior is consistent with a decrease of the carrier density with time. The final value of the conductance at the last day of measurement, as well as its scattering rate, is essentially equal to the one reported before for the 2D response in our study of the thickness dependence of the THz properties \cite{Rolando-Kerr}. We have shown previously that the spectral weight of this feature is completely accounted for by the expected carrier density of the topological surface states obtained from ARPES \cite{Hsieh09b}. Therefore, we assign the part that changes with time as the bulk response. Figure \ref{Fig1}(b) shows the imaginary part of the conductance. This also shows clearly the Lorentzian line shape due to the phonon, as well as the free electron response as a positive slope in the zero frequency limit. It is clear from these data that this slope decreases with time, indicating a loss of spectral weight with further exposure to ambient conditions. Data for a 128 QL thin film shown in fig. \ref{Fig1}(c) and (d) have the same qualitative behavior, and prove that the robustness of the 2D Drude response is an intrinsic feature of these topological insulator thin films.

In order to quantify the changes in the spectra, we fit them with a model ac conductivity function that effectively parametrizes the different components in the response. There are 2 Drude terms, one representing the bulk (D1) and the other the two surfaces of the film combined (D2), and there is a Drude-Lorentz (DL) oscillator for the response of the phonon:
\begin{eqnarray}
\nonumber
\sigma_{xx}(\omega)&=&\frac{\omega_{pD1}^2/4\pi}{i\omega-\Gamma_{D1}}+\frac{\omega_{pD2}^2/4\pi}{i\omega-\Gamma_{D2}}-\frac{i(\varepsilon_{\infty}-1)\omega}{4\pi}\\
&&-\frac{i\omega\omega_{pDL}^2/4\pi}{\omega_{DL}^2-\omega^2-i\omega\Gamma_{DL}}
\label{Drude}
\end{eqnarray}

\noindent Here $\varepsilon_{\infty}$ represents the high frequency optical transitions contributions to the dielectric function. $\omega_{pDi}$ and $\Gamma_{Di}$ are the Drude plasma frequency and scattering rate for the $i$th Drude component; $\omega_{DL}$, $\omega_{pDL}$ and $\Gamma_{DL}$ are, respectively, the center and plasma frequencies, and scattering rate of the optical phonon. We obtain the conductance by multiplying the ac conductivity by $t$ the film's thickness, $G_{xx} = \sigma_{xx}\times t$. We show in figure \ref{Fig2}(a) and (b) the results of the fits on the real part of the conductance with the different components plotted as individual traces for days 1 and 25, respectively. Fits for the other days are of similar quality. It is clear that the largest change between the 1st and 25th day of measurements occurs on the Drude component that we associate with the bulk response, and in particular it is the spectral weight ($\omega_p^2$), as opposed to the scattering rate, that is most affected. This Drude component corresponds to the bulk because we already know that the remaining contribution to the low frequency spectral weight after weeks of exposure behaves two-dimensionally, as was shown previously \cite{Rolando-Kerr}. In panels (c) and (d) we plot the imaginary conductances, where fits obtained simultaneously are of similar quality. Therefore, the results of the fit show that the bulk response changes significantly with time as the film is exposed to atmospheric conditions, whereas the surface remains mostly unaffected.

\begin{figure}[b]
\includegraphics[width=.9\columnwidth]{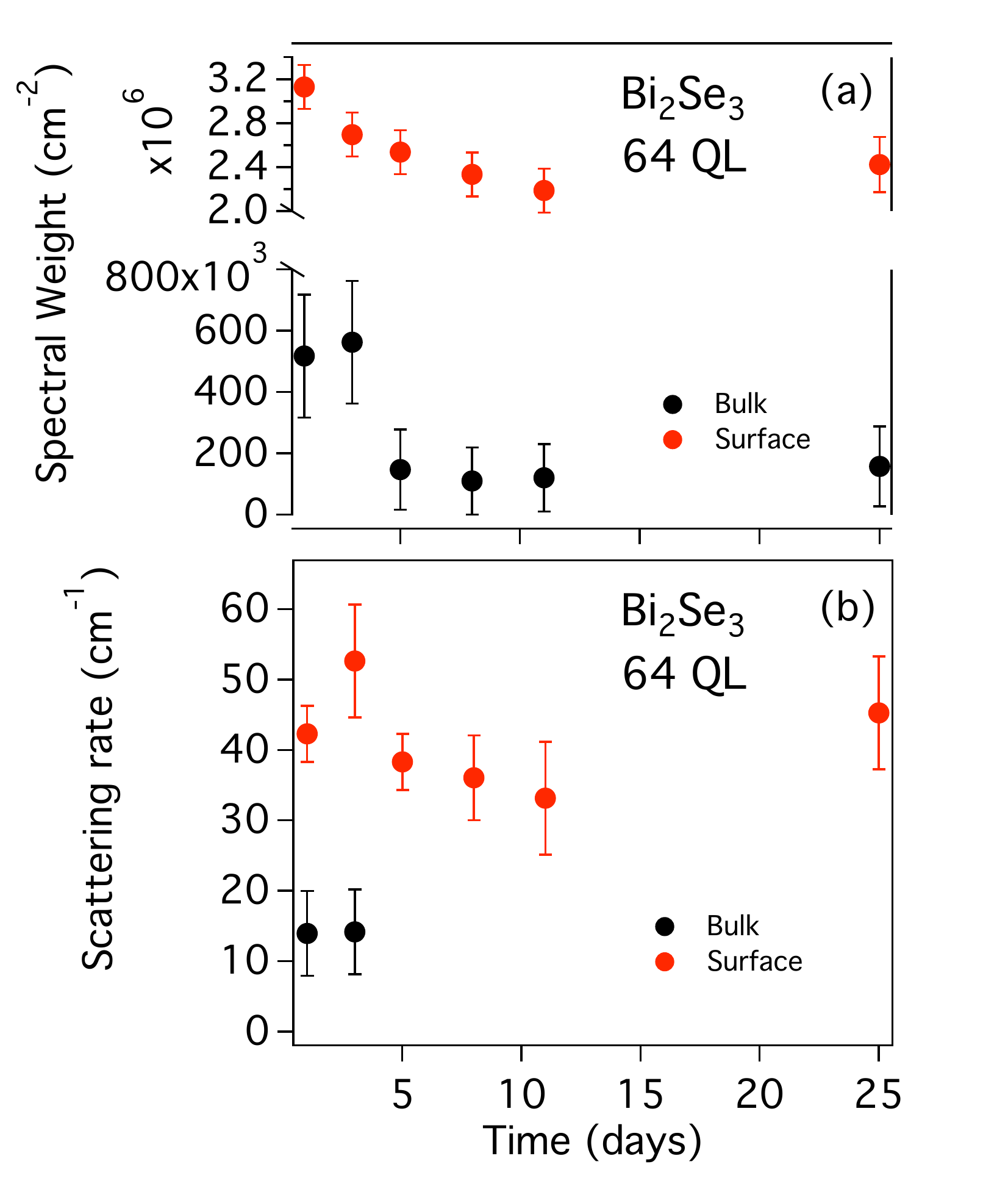}
\caption{(Color Online). \textbf{(a)} Spectral weights ($\frac{\omega_P^2}{4\pi^2}$) of bulk (black) and surface (red/gray) Drude components in a thin film of 64 QL as a function of measurement day. \textbf{(b)} Drude scattering rates for bulk and surface as a function of time for the same sample.}
\label{Fig3}
\end{figure}

In Figure \ref{Fig3}(a) and (b) we plot the time dependence of the fit parameters of the Drude components for the 64 QL sample. After the first 2 days, the Drude spectral weight of the bulk in the 64 QL film decreases rapidly and after day 10 stays relatively constant. We note, however, that after day 10 the spectral weight in this Drude component is so small that it is very difficult to distinguish it from zero, therefore the error bars on the plot reach zero on these cases. The Drude component of the surface, on the other hand, decreases more gradually in the first few days, but then seems to level off for the remainder of the experiment, staying constant within the accuracy of the measurement. The scattering rate (proportional to the mobilities) of the surfaces seems to stay mostly unchanged within this time scale, indicating that the carrier density is the property that changes the most with exposure. We note that at the lowest measured frequencies and temperatures, for films that have aged the most, the surface contribution to the conductance can be up to 80\% of the total conductance. This supports the scenario we proposed before \cite{Rolando-Kerr} that the THz response is dominated by the 2D carriers, and, surprisingly, allows us to conclude that exposing the films to atmospheric conditions is a novel way of enhancing the surface to bulk ratio to the total film's conductance.

The reduction of the bulk carrier concentration most probably results from the filling in of the Se vacancies present in the as-grown film by the oxygen in the atmosphere \cite{Brahlek-11a}. This effect would need to be stronger than the \textit{n}-doping effect that water molecules have on the surface\cite{Brahlek-11a,Benia11} in order to be consistent with our experiment. We would like to emphasize that the kind of aging effect we are reporting, which takes place over days in atmosphere conditions, is clearly different than the kind of aging that takes place over hours in UHV \cite{Bianchi10,King11}, which has been shown to raise the surface Fermi level.  The present case is probably driven by slow atomic diffusion into the bulk, while the UHV case is presumably driven by much faster physio- or chemisorption on the surface. We see no evidence for a third spectral feature that could derive from topologically trivial surface states.  One might speculate that these states, if they exist, have mobilities so low that they don't make an appreciable contribution to the conductance within our spectral window.

In conclusion, we have shown that exposure of thin films of the topological insulator Bi$_2$Se$_3$ to atmospheric conditions reduces the contribution of the bulk to the total film's conductance by significantly changing the carrier concentration. Using THz spectroscopy we were able to separate the effects on the mobility and carrier density of both the surface and bulk carriers, and thus conclude that in the time scales of days, for films exposed directly to the atmosphere, Se vacancies are replaced with Oxygen ions, which reduces the bulk carrier concentration without affecting the surface states properties that significantly. We think that the use THz spectroscopy will continue to aid in the study of the response of the topological surface states and other 2D electron systems.

This work was supported by DOE grant for The Institute of Quantum Matter at JHU DE-FG02-08ER46544 and by the Gordon and Betty Moore Foundation. The work at Rutgers was supported by IAMDN of Rutgers University, NSF DMR-0845464 and ONR N000140910749.

\bibliography{TopoIns}
\end{document}